\documentclass[
twocolumn,
superscriptaddress,
longbibliography,
pra,
tightenlines,
aps,
]{revtex4-1}
\pdfoutput=1

\usepackage[utf8]{inputenc}
\usepackage[T1]{fontenc}
\usepackage[english]{babel}
\usepackage[normalem]{ulem}
\usepackage{amsmath}
\usepackage{amssymb}
\usepackage{amsfonts}
\usepackage{mathrsfs}
\usepackage{dsfont}
\usepackage{bm} 
\usepackage{textgreek} 
\usepackage{dcolumn}
\usepackage{graphicx}
\usepackage[caption=false]{subfig}
\usepackage{enumerate}
\usepackage[shortlabels]{enumitem}
\usepackage{hyphenat}
\usepackage[bookmarks=false,colorlinks=true,linkcolor=blue,citecolor=blue,urlcolor=blue]{hyperref}
\usepackage[capitalise]{cleveref}
\crefname{section}{Sec.}{Secs.}
\Crefname{section}{Section}{Sections}
\usepackage{hypcap} 
\usepackage{floatrow}
\usepackage{tablefootnote}
\usepackage{array}
\usepackage{braket}
\usepackage{siunitx}

\newcommand{\Tr}{\text{Tr}}

\begin{document}

\title{Towards heralded distribution of polarization entanglement}

\author{F.~Joseph~Marcellino}
\affiliation{Department of Applied Physics, University of Geneva, CH-1211 Geneva, Switzerland}

\author{Patrik~Caspar}
\affiliation{Department of Applied Physics, University of Geneva, CH-1211 Geneva, Switzerland}

\author{Tiff~Brydges}
\affiliation{Department of Applied Physics, University of Geneva, CH-1211 Geneva, Switzerland}

\author{Hugo~Zbinden}
\affiliation{Department of Applied Physics, University of Geneva, CH-1211 Geneva, Switzerland}

\author{Rob~Thew}
\email[Corresponding author: ]{Robert.Thew@unige.ch}
\affiliation{Department of Applied Physics, University of Geneva, CH-1211 Geneva, Switzerland}

\date{\today}

\begin{abstract}
Distributing entangled states over potentially long distances provides a key resource for many protocols in quantum communication and quantum cryptography. Ideally, this should be implemented in a heralded manner. By starting with four single-photon states, we cascade two single-photon path-entangled states, coded in orthogonal polarizations, to distribute and herald polarization entanglement in a single quantum repeater link architecture. By tuning the input states to minimize (local) losses, the theoretically achievable fidelity to the target state without postselection approaches 1, while sacrificing heralding rates. We achieve a fidelity to the target state of over 95\% after postselection, providing a benchmark for the experimental control. We show that the fidelity of the heralded state without postselection scales predictably and also identify various practical challenges and error sources specific to this architecture, and model their effects on the generated state. While our experiment uses probabilistic photon-pair sources based on spontaneous parametric down-conversion, many of these problems are also relevant for variants employing deterministic photon sources. 

\end{abstract}

\maketitle

\section{Introduction}
\label{sec:introduction}

\begin{figure}
\capstart
\includegraphics[width = 0.95 \textwidth]{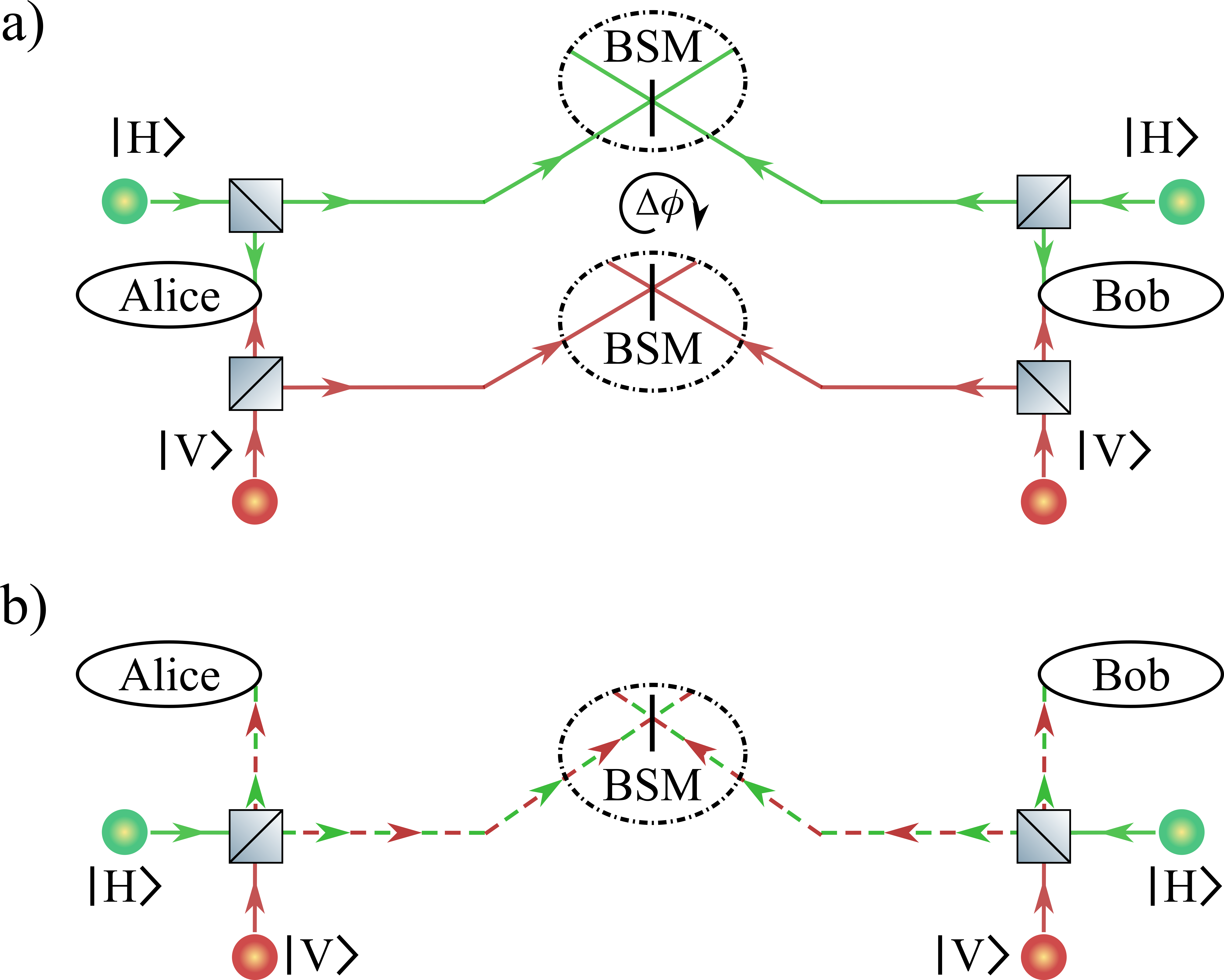}
\caption{a) Schematic of dual-rail single-photon path entanglement distribution. b) Schematic of multiplexed cascaded entanglement distribution. Grey squares indicate variable beam splitters}
\label{fig:concept}
\end{figure}

The distribution of entanglement represents a fundamental challenge for future quantum communication networks. Perhaps the two most important primitives are quantum repeaters and device-independent quantum key distribution (DIQKD). One of the most promising quantum repeater architectures relies on single-photon entangled states~\cite{Duan2001,Sangouard2007}, which have a definite scaling advantage but require phase stabilization in network links. In these schemes, the single-photon entanglement is used to build up polarization-entangled states in a dual-rail configuration, as shown in Fig.~\hyperref[fig:concept]{\ref{fig:concept}(a)} \cite{Sangouard2011}. For this to work, the phase in this distributed interferometer, $\Delta\phi$, needs to remain stable while the entanglement is generated in the two modes, which can be challenging to implement over large distances. 

In practice, however, we can simplify this, at least partially, as in Fig.~\hyperref[fig:concept]{\ref{fig:concept}(b)}. Here, the two orthogonal polarization modes $\{H, V\}$ are multiplexed into a single optical fiber. Now one only needs to ensure that the phase fluctuations~\cite{Minar2008} are stabilized on a timescale comparable with the generation rate in the link, i.e., the time difference between the $H$ and $V$ photons propagating in the same fiber link. This is analogous to the time difference between the temporal modes of time-bin qubits, which have proven their robustness to phase fluctuations on quantum memory timescales~\cite{Ortu22}.

We call such schemes cascaded entanglement swapping as it can be seen to be built up of two separate and independent single-photon entanglement swapping operations that are cascaded together - the central Bell State Measurement (BSM) erasing information about the origin (A or B) of the detected photon and hence which source emitted the photon originally. If we consider that we rotate the polarization of the states before the BSM, one can also see the similarities with the concept of type II fusion operations~\cite{Browne2005, Zhang2008}, albeit used for a distributed fiber-based architecture here. In the typical quantum repeater scenario where $A$ and $B$ have quantum memories, the arrival time of the photon at the BSM and hence the heralding signal for the ``H” and ``V” channels can also take place at different times, provided that the memories are eventually read out in such a way as to again erase this timing information.

This architecture is isomorphic with a recent proposal for DIQKD that exploits single photon sources to generate the distributed entanglement~\cite{Kolodynski2020}. Central to both of these primitives is that the architecture sacrifices loss for the photons sent to the BSM to minimize loss for the final distributed, and heralded, state of Alice and Bob. This is achieved by varying the beam splitter ratios, (grey squares) indicated in Fig.~\hyperref[fig:concept]{\ref{fig:concept}(b)}, such that the transmission of photons going to Alice and Bob tends towards 1. This is critical for both primitives but essential for the detection loophole-free violation of Bell inequalities~\cite{Bell1964,Brunner2014,Hensen2015,Giustina2015,Shalm2015} needed for DIQKD. Recent theoretical work has considered some experimentally relevant challenges to achieve this in practice, but this has been mostly related to the implementation of deterministic single photon sources~\cite{Gonzalez2022,Gonzalez2023}.

In this paper, we focus on the challenges facing the generation and distribution of entanglement using such schemes. We first recall the theoretical basis for these primitives before introducing the experimental setup and characterizing the photons in terms of their spectral purity and indistinguishability. We then measure the postselected fidelity of the distributed entangled states to provide a benchmark, before tuning the state towards the limiting case required for quantum repeaters and DIQKD. Finally, we discuss some practical constraints and their implications for distributing high fidelity heralded entanglement with both probabilistic and deterministic photon sources.

\section{Concept}
\label{sec:concept}

In the ideal case, the input state for the scheme consists of four linearly polarized photons
\begin{equation}
    \begin{split}
        \ket{\Psi}_\mathrm{in} = &(\alpha\ket{H}_{A} + \beta\ket{V}_{A})\otimes(\alpha\ket{V}_{A'} + \beta\ket{H}_{A'})\otimes \\
        & (\alpha\ket{H}_{B} + \beta\ket{V}_{B})\otimes(\alpha\ket{V}_{B'} + \beta\ket{H}_{B'})
    \end{split}
\end{equation}
where subscripts $A$ $(B)$ and $A'$ $(B')$ indicate the two incoming path modes on Alice's (Bob's) side, and $|\beta|^2 = 1 - |\alpha|^2$ define the variable beam splitter (VBS) transmission and reflection settings for the polarization modes (see Fig.~\hyperref[fig:concept]{\ref{fig:concept}}).

In practice, as we see in Fig.~\ref{fig:setup}, the VBS are replaced by polarizing beam splitters (PBS) and wave plates. The two photons on Alice's side are interfered on one PBS, and the two on Bob's side on another PBS, such that one output mode of each PBS is distributed to Alice/Bob and the other is sent to a central heralding station, where a BSM is performed (see Fig.~\ref{fig:setup}). Here, the two incoming photons are rotated by $45^\circ$ and interfered on a $50/50$ beam splitter, the outputs of which pass through additional PBSs and are detected. Upon detection of exactly one horizontal photon in one output arm of the beam splitter and one vertical photon in the other arm, denoted path modes $e$ and $f$ respectively, the two remaining photons are projected into the Bell state $\ket{\Psi^-}_{ab} = \frac{1}{\sqrt{2}}(\ket{H,V}_{ab} - \ket{V,H}_{ab})$. 

\begin{figure}
\centering
\includegraphics[width = 0.95\textwidth]{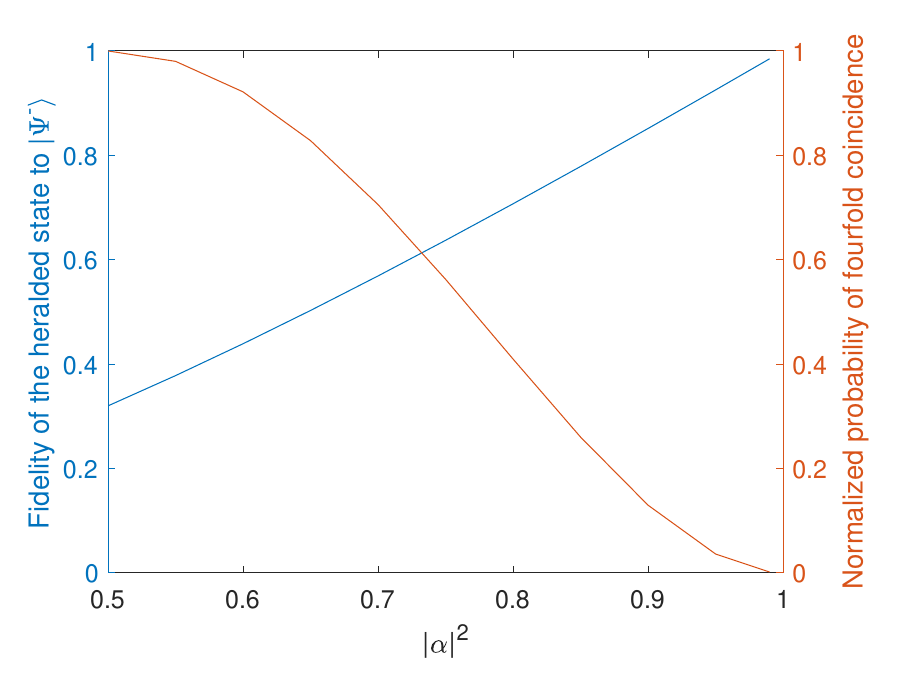}
\caption{\label{fig:fig_theory_curve.pdf} Fidelity of heralded state (blue) to $\ket{\Psi^-}$ and normalized fourfold coincidence probability (red) as a function of the variable beam splitter transmission, $|\alpha|^2$, for an idealized case.}
\end{figure}
The global state before detection has the form 
\begin{equation}
    \begin{split}
        \ket{\Psi} = &(N_0\alpha^2\beta^2(\ket{H,V}_{ef} + \ket{V,H}_{ef})\ket{\Psi^-}_{ab} \\
        & + N_1\alpha\beta^3\ket{\psi_3}_{ef}\ket{\psi_1}_{ab} \\
        & + N_2\beta^4\ket{\psi_4}_{ef}\ket{0}_{ab} + (...)
    \end{split}
\end{equation}
where $\ket{\psi_n}_{pq}$ denotes the portion of the global state with $n$ photons in path modes $pq$, $(...)$ contains all terms with one or fewer photons in the Bell state measurement path modes, and $N_j$ are combinations of normalization coefficients. Coincidences between $He$ ($Ve$) and $Vf$ ($Hf$) detectors thus herald $\ket{\psi_1}_{ab}$ and $\ket{0}_{ab}$ in addition to the desired $\ket{\Psi^-}_{ab}$. 

This problem can be mitigated by using photon number resolving (PNR) detectors \cite{Mattioli2015}, which allow one to discriminate between $\ket{H,V}_{ef}$ or $\ket{V,H}_{ef}$, which herald the desired state, and $\ket{\psi_3}_{ef}$ or $\ket{\psi_4}_{ef}$, which do not, thus improving the heralding efficiency for the desired state. Similarly, we see that by adjusting the polarization of the input photons, one can exploit the scaling in $|\alpha|^2$ of the desired term, such that the probability of heralding $\ket{\Psi^-}$ approaches $1$ in the ideal case (Fig.~\ref{fig:fig_theory_curve.pdf}). Here, we also see that this advantage comes at the cost of a significant reduction in the probability of a heralding event. In this experiment, probabilistic photon-pair sources based on spontaneous parametric down-conversion (SPDC) are used, however, most results are also relevant for deterministic sources. The fidelity and scaling of both the heralded and postselected states, including effects due to SPDC, are experimentally explored and theoretically modeled in the Discussion section. 

\section{Experiment}
\label{sec:experiment}

\begin{figure*}[!t]
\capstart
\includegraphics[width = 0.8 \textwidth]{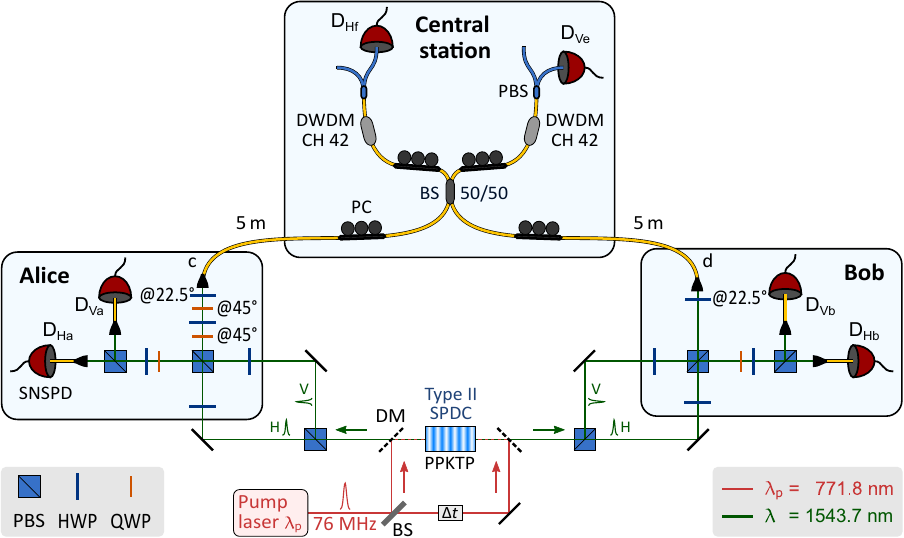}
\caption{\label{fig:setup} Setup for the distribution of polarization entanglement. The pulsed pump laser bidirectionally pumps the nonlinear SPDC crystal to create degenerate photon pairs, which are separated on each side by a PBS. The four photons are then recombined on subsequent PBSs in linear polarization states. One part of the state is kept locally, whereas the other part of the state from each side is sent to the Bell State Measurement at the central station where we detect coincidences between detectors $\mathrm{D}_{Ve}$ and $\mathrm{D}_{Hf}$. 
BS, beam splitter; DM, dichroic mirror; DWDM, dense wavelength division multiplexer; HWP, half-wave plate; PBS, polarizing beam splitter; PC, polarization controller; PPKTP, periodically poled potassium titanyl phosphate; QWP, quarter-wave plate; SNSPD, superconducting nanowire single-photon detector; SPDC, spontaneous parametric down-conversion.}
\end{figure*}

The experimental implementation of our scheme is shown in Fig.~\ref{fig:setup}. 
We use a periodically poled potassium titanyl phosphate (PPKTP) nonlinear crystal which is bidirectionally pumped by a picosecond pulsed laser (\SI{76}{MHz} repetition rate) at $\lambda_p=\SI{771.85}{nm}$ to create degenerate photons at $\lambda_s=\lambda_i=\SI{1543.7}{nm}$. For both pump directions, the photon pair creation probability per pump pulse is around $p\approx \num{3E-3}$ in order to keep the probability of having double-pair emissions low. The orthogonal signal and idler modes of each pair are separated after their generation at the first PBS. After that, we prepare the initial linearly polarized  states by means of half-wave plates (HWP) before they are combined on PBSs. There is a walk-off between the two orthogonally polarized photons of around \SI{4.5}{ps} due to the birefringence of the PPKTP crystal, which is compensated for by positioning the combining PBS to set a \SI{1.35}{mm} shorter path for the $H$ photons than for the $V$ photons.

In the distributed path modes on Alice's and Bob's sides, we have a quarter-wave plate (QWP) and a HWP followed by a PBS and two superconducting nanowire single-photon detectors (SNSPDs) on each output mode in order to be able to measure the state in any basis. The other two modes are coupled into fiber and combined on a 50/50 fiber BS at the central station. The three wave plates on Alice's side before coupling into the fiber allow for relative phase control between the modes $Hc$ and $Vc$, which suffices to adjust the phase $\phi$ in the desired part of the heralded state $(\ket{H,V}_{ab}+e^{\mathrm{i}\phi}\ket{V,H}_{ab})$. Additional HWPs on both sides enact the $45\deg$ rotation required to prevent two orthogonally-polarized photons from the same side from causing heralds. The polarization controllers (PC) are used for polarization alignment by sending light in the reverse direction and measuring the polarization state with a polarimeter in free space. After the BS at the central station, we use \SI{200}{GHz} DWDM at ITU channel 42 (\SI{1543.73}{nm}) to remove the spectral sidebands in order to herald photons on Alice's and Bob's sides with high spectral purity. We then employ fiber PBSs to project on the horizontal and vertical polarization modes before detecting photons with SNSPDs. Here, we only use two SNSPDs at the central station, one in path  mode $Ve$ and one in $Hf$. 

All SNSPDs used are MoSi single-meander detectors \cite{Caloz:2018,Autebert:2020} with detection efficiencies between \SI{79}{\%} and \SI{90}{\%}, noise counts below 300\,$s^{-1}$ and detection event windows of 1\,ns. The symmetric coupling efficiency is measured to be around \SI{90}{\%}, whereas the transmission efficiencies through the optical elements are between \SI{72}{\%} and \SI{83}{\%}, depending on the path. When detecting directly after the fiber couplers in the spatial modes $c$ and $d$, without going through the BSM setup, the heralding efficiencies for all modes are between \SI{65}{\%} and \SI{75}{\%} (this includes coupling and transmission efficiencies, but excludes detection efficiencies). We achieve a fourfold coincidence rate of approximately 0.5 events $s^{-1}$  at $|\alpha|^2 = 0.5$, which scales predictably at higher values of $|\alpha|^2$ (see Fig.~\ref{fig:fig_theory_curve.pdf}).

\begin{figure}
\centering
\includegraphics[width = 0.95\textwidth]{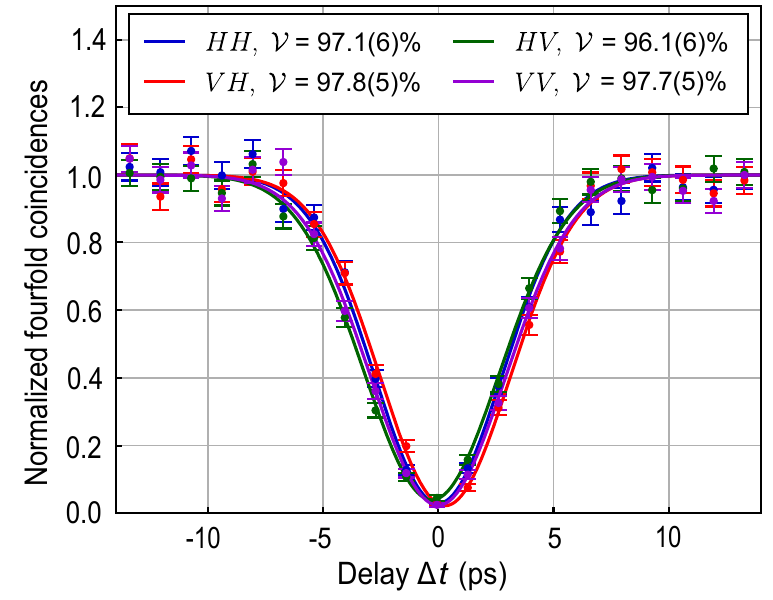}
\caption{\label{fig:hom} Hong-Ou-Mandel (HOM) interference for all four combinations of photons ($HH$, $HV$, $VH$ and $VV$) from Alice's and Bob's side sent to the 50/50 BS at the Bell state measurement, 50s of measurement. The 4-fold coincidences between the two local heralding detectors and the two detectors at the central station are recorded as a function of the delay $\Delta t$ in Alice's pump path. The visibilities $\mathcal{V}$ are calculated for all four HOM dips from a weighted inverse Gaussian fit, where the weight of each data point is $1/\sigma$ under the assumption of Poissonian count statistics with standard deviation $\sigma$.}
\end{figure}

The impact of finite indistinguishability and spectral purity of the generated photons is quantified by measuring their Hong-Ou-Mandel (HOM) interference, as shown in Fig.~\ref{fig:hom}. We send all four different combinations of horizontally/vertically polarized photons on Alice's and Bob's side to the 50/50 BS at the Bell state measurement, with the other photon of each pair detected locally.
The wave plates on Alice's side before coupling into fiber are used to ensure that the polarization state of Alice's photons matches the one of Bob's photons at the 50/50 BS. The 4-fold coincidences between Alice's and Bob's detectors and the two detectors at the central station (in the modes $He$ and $Hf$ for $HH$ and $VH$ photons, or in the modes $Ve$ and $Vf$ for $HV$ and $VV$ photons) are recorded as a function of the delay $\Delta t$ between Alice and Bob's pair generation events. 
We observe HOM visibilities of around \SI{97}{\%} for all four combinations. Moreover, we see that the photons from different input paths are temporally well aligned as all four HOM dips are centered around a delay of $\Delta t=\SI{0}{ps}$. 

\section{Results}
\label{sec:results}

\begin{figure}[t!]
\centering
\includegraphics[width = 0.95\textwidth]{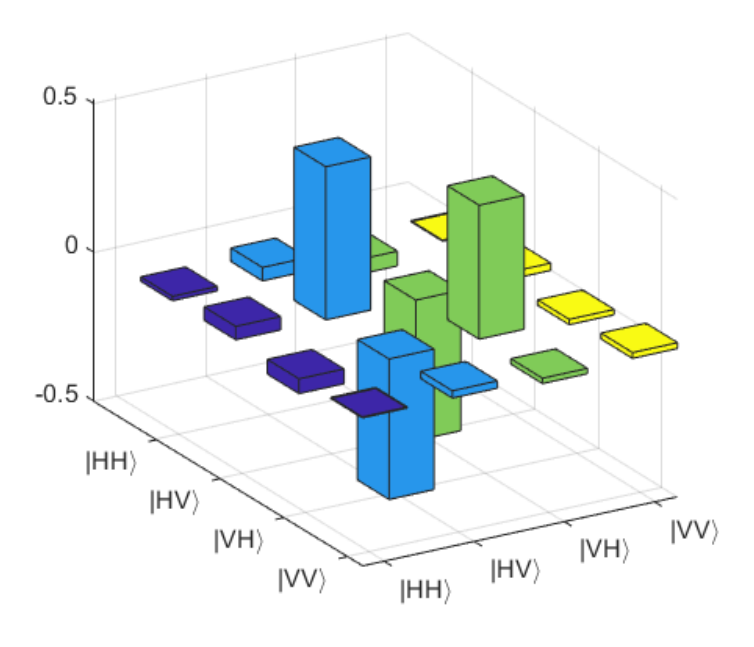}
\caption{\label{fig:rho} Real part of the density matrix obtained from a quantum state tomography measurement. All elements of the imaginary part (not shown) have absolute values $< 0.015$.}
\end{figure}

We perform quantum state tomography on the distributed part of the fourfold postselected state, i.e., the state conditioned on detecting one photon in the \textit{Hf} mode, one in the \textit{Ve} mode, and one in each of the \textit{Ha}/\textit{Va} and \textit{Hb}/\textit{Vb} modes. The real part of the reconstructed two-qubit density matrix $\rho$ is given in Fig.~\ref{fig:rho} (all elements of the imaginary part have absolute values $< 0.015$). The fidelity of the postselected state to the target $\ket{\Psi^{-}}$ state $F \equiv \bra{\Psi^{-}}\rho \ket{\Psi^{-}}$ (referred to henceforth simply as ``fidelity") is computed as $95.51\pm 0.08\,\%$, and the two-qubit state purity $\Tr(\rho^2)$ as $92.71\,\%$. We estimate the error in $F$ due to finite statistics by sampling random density matrices using a Metropolis-Hastings algorithm according to their propensity to have produced the observed statistics, then generating a probability density function of the associated fidelities (For more detail about this method, see \cite{Faist2016}). The non-unit state purity can be attributed primarily to imperfect spectral purity and temporal indistinguishability of the input photons, as well as multi-pair events; these introduce additional frequency, temporal, or photon number modes that take the state outside the qubit space, which are then effectively traced over by the detection process, resulting in a mixed state.

We perform two additional measurements, setting $|\alpha|^2$ on each side to 0.66, then 0.75, resulting in a respective 2:1 and 3:1 bias towards Alice and Bob's detectors. As mentioned, increasing $|\alpha|^2$ increases the fidelity of the heralded (non-postselected) state at the expense of decreased rates. To compute the fidelity of the heralded state to $\ket{\Psi^-}$, we first compute the ratio of total fourfold events to total heralding events, then weight the result by the measured fidelity of the fourfold postselected state (in doing so we assume that any state outside the qubit space, e.g., $\ket{H}_a\otimes\ket{0}_b$, is orthogonal to $\ket{\Psi^-}$). The fidelity of the heralded state as a function of $|\alpha|^2$ is given in Fig.~\ref{fig:fid_vs_imb_meas}. Of note is the slight decrease in fidelity between the 2:1 and 3:1 cases, rather than an increase; we explore reasons for this phenomenon in the following.
\section{Discussion}
\label{sec:discussion}

The fidelity of the heralded state is determined by several effects, most importantly: unwanted terms in the initial state due to the SPDC process, e.g. double pair contributions, losses for the distributed modes, detector efficiencies, and pollution of the polarization states by experimental imperfections, such as finite extinction ratios on PBSs and misalignment of the bases the PBSs define. In the following, we elaborate on each of these effects individually. 

\begin{figure}[t!]
    \capstart
    \includegraphics[width = .95\textwidth]{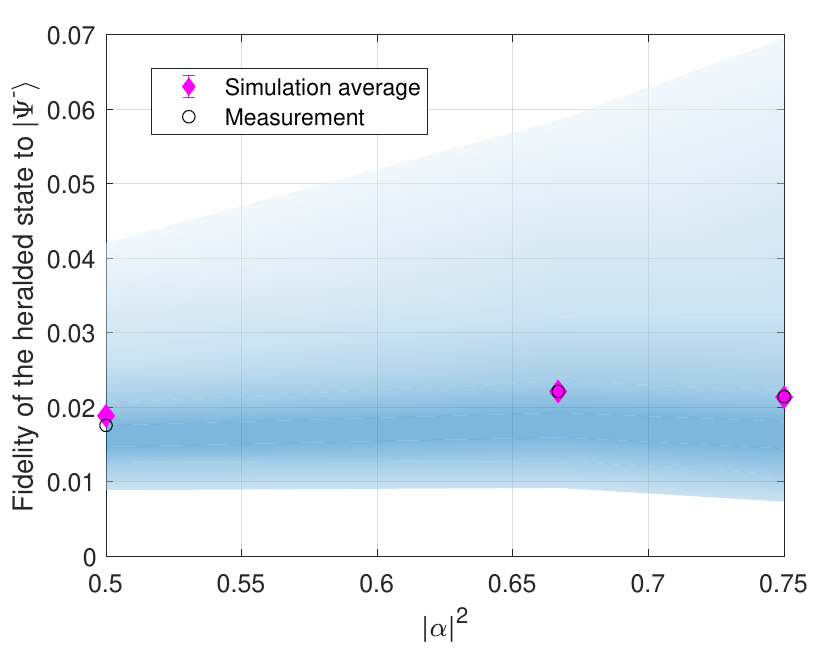}
    \caption{Fidelity of heralded state to $\ket{\Psi^-}$, measured values (open black circles) and simulation (blue region), including effects from PBS leakage/misalignment and detector loss. The simulation was run with 1000 sets of randomly chosen phases added to the initial TMSV state and during preparation of the states before swapping; the range of outcomes is indicated by the blue region, with darker color indicating higher probability. The average fidelity over these phases is indicated in magenta.}
    \label{fig:fid_vs_imb_meas}
\end{figure}
First, let us consider the source errors associated with multiphoton ($g^{(2)}(0)\neq 0$) events. The initial state after the crystal on each side is a two-mode squeezed vacuum (TMSV) state with squeezing parameter $\lambda$:
\begin{equation}
    \ket{\psi} = \cosh^{-1}(\lambda)\sum_{n=0}^{\infty}e^{in\theta}\tanh^n(\lambda)\ket{n,n}
\end{equation}
which, for small $\lambda$, can be approximated as 
\begin{equation}\label{Eq:spdcState}
\begin{split}
    & \ket{\psi}\approx  \frac{1}{\sqrt{1+p^2+p^4}}\Big(\ket{0,0}+ e^{i\theta}p\ket{1,1} \\
   & \hspace{1cm} + e^{2i\theta}p^2\ket{2,2} + O(p^3)\Big), 
\end{split}
\end{equation}
where $p$ gives the probability per pulse of down-converting a pump photon, and is proportional to the pump power.

The initial state is then the tensor product of the TMSV states on Alice and Bob's sides, respectively:
\begin{equation}\label{Eq:InitialState}
    \ket{\Psi_\mathrm{init}} = \ket{\psi}_{AA'} \otimes \ket{\psi}_{BB'}.
\end{equation}
Successful execution of the protocol necessitates the input state $ \ket{1,1}_{AA'} \otimes \ket{1,1}_{BB'}$, corresponding to a single pair of photons generated on each side -- that is, a single photon at each of the four inputs. As we see from Eq.~\eqref{Eq:spdcState} and Eq.~\eqref{Eq:InitialState}, the first limitation of the fidelity is due to the contributions from a double pair on one side and vacuum on the other, such that the desired state is only heralded one third of the time \cite{Takeoka2015}. 

\begin{figure}[t!]
    \capstart
    \includegraphics[width = .95\textwidth]{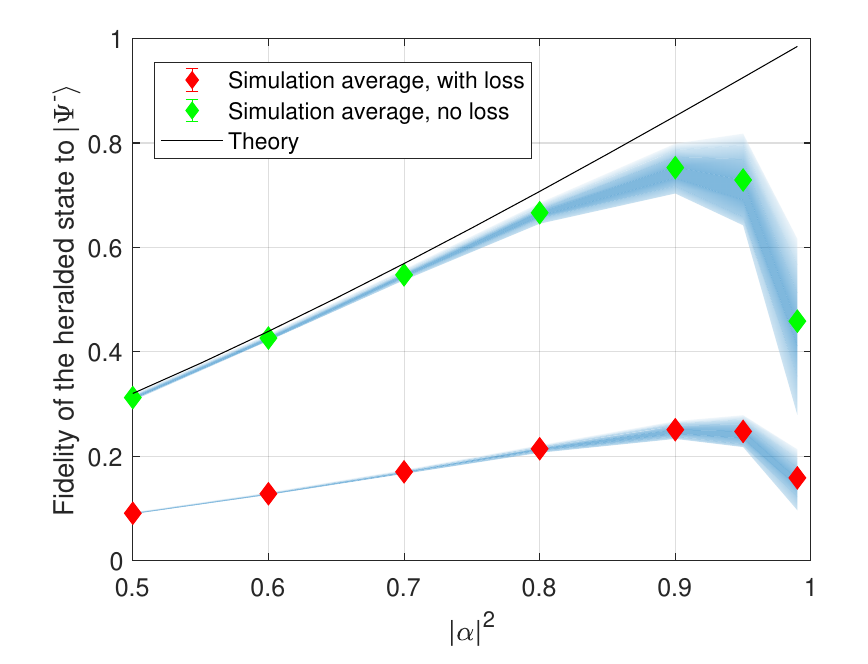}
    \caption{Simulated fidelity of heralded state to $\ket{\Psi^-}$, for the ideal input state $\ket{1,1}_{A,A'} \otimes \ket{1,1}_{B,B'}$, including effects from PBS leakage and misalignment. The simulations were run with 1000 sets of randomly chosen phases added during preparation of the states before swapping; the range of outcomes are indicated by the blue regions, with darker color indicating higher probability. The average fidelities over these phases in the cases with and without detector loss are indicated in red and green respectively. }
    \label{fig:fid_vs_imb_dots}
\end{figure}
Vacuum contributions, due to loss or non-unit detector efficiencies on Alice and Bob's sides also reduce the fidelity of the heralded state. This loss is of critical importance for the scaling of quantum repeaters, or overcoming detection loophole thresholds for DIQKD, and motivates the current architecture of putting all the transmission distance, and loss on the BSM channels. Detector noise at the BSM can also result in heralding of unwanted states, although, in practice, for low dark-count detectors the probability of a coincidence between a dark count and a real photon, or between two dark counts, is relatively low~\cite{Sekatski2012} given the detector noise of the SNSPDs. 

Next, we explore experimental imperfections related to the polarization encoding. The effects of PBS leakage and misalignment on the final state are complex, but we illustrate with a simple example where leakage causes a perfect $\ket{1,1}_{A,A'} \otimes \ket{0,0}_{B,B'}$ input state to result in a heralding event.

We first consider an ideal PBS with input modes $1$ and $2$ and output modes $3$ and $4$. This is a simple routing device that performs the transformation
\begin{equation}
    (\gamma h^{\dagger}_{1} + \delta v^{\dagger}_{1})\ket{0} \rightarrow (\gamma h^{\dagger}_{3} + \delta v^{\dagger}_{4})\ket{0}   
\end{equation}
where e.g., $h^{\dagger}_{1}$ creates a horizontally polarized photon in path mode 1, $\ket{0}$ is the vacuum, and $|\gamma|^2 + |\delta|^2 = 1$. In contrast, physical PBSs suffer from finite extinction ratios (``leakage”), which result in a small amount of photons with the wrong polarization in each path mode:
\begin{equation}\label{leak}
\begin{split}
    & (\gamma h^{\dagger}_{1} + \delta v^{\dagger}_{1})\ket{0} \rightarrow (\gamma((1-p_{\mathrm{leak},h}) h^{\dagger}_{3} + p_{\mathrm{leak},h} h^{\dagger}_{4}) \\
    & \hspace{2cm} + \delta((1-p_{\mathrm{leak},v}) v^{\dagger}_{4} -p_{\mathrm{leak},v} v^{\dagger}_{3}))\ket{0}   
\end{split}
\end{equation}
where the beam splitter phase shift is applied arbitrarily to the $v^{\dagger}_{3}$ term, and we neglect insertion loss. 

We now follow the evolution of a $\ket{1,1}_{A,A'}$ input state on Alice's side. Assuming a perfect splitting PBS, setting $|\alpha|^2$ to $0.5$, and denoting the transmitted and reflected path modes after the splitting PBS with subscripts A and A' respectively, we can write the state just before the combining PBS as 
\begin{equation}
    \frac{1}{2}(h^{\dagger}_{A}h^{\dagger}_{A'} + h^{\dagger}_Av^{\dagger}_{A'} + v^{\dagger}_Ah^{\dagger}_{A'} + v^{\dagger}_Av^{\dagger}_{A'})\ket{0}.
\end{equation}
Suppose first that the combining PBS is also ideal. Denoting the distributed and central BSM path modes with subscript $a$ and $c$ respectively, the state is then transformed into
\begin{equation}\label{eq:state_post_comb_pbs}
    \frac{1}{2}(h^{\dagger}_ch^{\dagger}_a + h^{\dagger}_cv^{\dagger}_c + v^{\dagger}_ah^{\dagger}_a + v^{\dagger}_av^{\dagger}_c)\ket{0}.
\end{equation}

Recalling that a heralding event requires one horizontal and one vertical photon in the BSM path mode, we see that the $h^{\dagger}_cv^{\dagger}_c$ term in Eq.~\eqref{eq:state_post_comb_pbs} can \textit{a priori} result in a heralding event; however, the effect of the additional HWP at $22.5^{\circ}$ in the BSM path is to rotate this term to $\frac{1}{2}(h^{\dagger2}_c - v^{\dagger2}_c)$, which cannot cause heralding events.

However, suppose now that the combining PBS suffers from leakage. The effect of this is to introduce terms of the form $h^{\dagger2}_c$ and $v^{\dagger2}_c$ to Eq.~\eqref{eq:state_post_comb_pbs}, which are then rotated by this additional HWP to $\frac{1}{2}(h^{\dagger2}_c \pm 2h^{\dagger}_cv^{\dagger}_c \pm v^{\dagger2}_c)$, which can result in heralding events. Moreover, as single-sided pair generations are more probable than $\ket{1,1}_{AA'} \otimes \ket{1,1}_{BB'}$ by a factor of $p$, this failure mode remains an important source of fidelity loss even granting high extinction ratio PBSs.

A similar effect is obtained from misalignment of either the splitting or combining PBS. The spatial orientation of a PBS defines the basis of the output polarization modes. If this axis is misaligned by $\phi$ with respect to the basis of the input modes, then the ideal transformation written in the input mode basis is instead
\begin{equation}
\begin{split}
    & (\gamma h^{\dagger}_{1} + \delta v^{\dagger}_{1})\ket{0} \rightarrow \big(\gamma(\cos{(\phi)} h^{\dagger}_{3} + \sin{(\phi)} h^{\dagger}_{4}) \\
    & \hspace{2.8cm} +\delta(\cos{(\phi)} v^{\dagger}_{4} -\sin{(\phi)} v^{\dagger}_{3})\big)\ket{0} 
\end{split}  
\end{equation}
 resulting in unintended polarization modes in both path modes.

We simulate the experiment, including the effects previously discussed, using MATLAB. We model detector losses as a beam splitter operation between the desired mode, $a^{\dagger}$, and a loss mode, $x^{\dagger}$,
\begin{equation}
    a^{\dagger} \rightarrow \sqrt{\eta} a^{\dagger} + \sqrt{1-\eta}x^{\dagger},
\end{equation}
where $\eta$ is the transmission. We use measured values for imperfections, consisting of previously discussed coupling values, PBS extinction ratios between $20$ and $40$\,dB, and splitting PBS misalignment of between $1$ and $2$ degrees. 

An additional effect of PBS leakage and misalignment is to introduce phase dependence to the protocol performance. Each term $\ket{n,n}$ in the TMSV state carries a phase $e^{ni\theta_j}$, where $\theta_j$ is determined by the pump phase and may vary in time and between Alice and Bob's sides. Ideally, the pump phases manifest only as a global phase on the targeted final state; however, phases carried by unwanted multiphoton terms can reduce the fidelity of the heralded state. Photons traveling in different path modes on Alice's and Bob's sides can also accrue different phases, resulting in a relative phase between the $\ket{HV}$ and $\ket{VH}$ terms of the final $\ket{\Psi^-}$ state, although this is corrected for by using a combination of quarter- and half-wave plates. 

As noted, PBS leakage and misalignment introduces unwanted terms to the states before swapping. The phase differences between these various unwanted terms can cause them to interfere constructively or destructively to affect the fidelity of the heralded state. Importantly, we note that, despite these phases drifting locally for Alice and Bob, we are not adversely affected by phase fluctuations over the fiber link - although not long, a fiber interferometer as depicted in Fig.~\hyperref[fig:concept]{\ref{fig:concept}(a)} would fluctuate so rapidly that all entanglement would be lost. By multiplexing the two polarization modes, this phase has minimal impact, especially in this case where there is no temporal delay between the two modes. Rather, the phases we are concerned with affect the state prior to coupling into fiber.

To understand the effects of these phases, we simulate the experiment adding 1000 randomly chosen sets of phases and indicate the range of corresponding fidelities, as well as the average at each value of $|\alpha|^2$, in Fig.~\ref{fig:fid_vs_imb_meas} (blue shaded region). In principle the phases could be stabilized such that unwanted terms maximally destructively interfere; this would amount to taking the upper bound of the shaded region. The model average (filled magenta circles) matches the measured fidelity (open black circles) to within 7\,\% for the 1:1 ratio, and within 1\,\% for 2:1 and 3:1. Fluctuations in the phases over the course of each measurement due to temperature or acoustic noise naturally result in a degree of averaging over the shaded region. Higher values of $|\alpha|^2$ lead to lower heralding rates and so require longer measurements to achieve the same statistics; longer measurements include more fluctuations, hence more averaging and better agreement with the overall model average. 

Input states much more closely approximating $\ket{1,1}_{AA'} \otimes \ket{1,1}_{BB'}$ are achievable with the use of deterministic single photon sources, such as those provided by semiconductor quantum dots~\cite{Arakawa2020} or multiplexed probabilistic sources \cite{Meyer-Scott2020}. We simulate our experiment, with and without detector loss, for an ideal input state; the results, as well as the theoretical bound for an ideal input state with no experimental imperfections, are given in Fig.~\ref{fig:fid_vs_imb_dots}. We observe that in this ideal input case, with no imperfections, the fidelity tends to 1 as $|\alpha|^2$ increases (black line). When imperfections are included the behavior changes. When taking into account beam splitter leakage without loss, we see there is an increase in fidelity as $|\alpha|^2$ grows, followed by a decrease at very high $|\alpha|^2$ as the amount of light sent intentionally to the BSM becomes comparable to the amount present due to leakage (green points). When subsequently including loss, we see that the effect persists; however, the overall fidelity is also reduced (red points). This behavior is of particular importance as it is not due to effects inherent to probabilistic SPDC sources, and so will impact implementations of the protocol which use deterministic single photon sources, ultimately limiting the achievable fidelity.


\section{Conclusion}
\label{sec:conclusion}

This scheme represents a promising approach for the preparation and distribution of high-fidelity heralded entangled states in future quantum networks. We demonstrated a postselected fidelity for the distributed state of over 95\,\% starting from 4 indistinguishable single photons and using simple linear optic circuits. This benchmark value for the fidelity indicates good control over the experimental conditions and enables us to confidently study the non-postselected regime. While our fidelity in the non-postselected regime is low, we highlight some of the effects degrading the performance of such protocols, such as using input states with $g^{(2)}\neq 0$. Crucially, we also show that our low non-postselected fidelity is not merely due to using probabilistic sources or having a non-zero $g^{(2)}$, but is inherently limited by real-world experimental imperfections beyond just loss, such as PBS leakage -- effects which will both impact the protocol even when implemented with ideal single photon sources. We show that the effect of the leaking PBSs is to fundamentally change the behavior of the non-postselected fidelity as a function of $|\alpha|^2$, with a very large $|\alpha|^2$ actually yielding both low rates and low non-postselected fidelities. 

Looking towards further improvements, PNR detection has recently started to achieve performance levels that can help improve heralded fidelities~\cite{Stasi2023, Resta2023} by effectively reducing the $g^{(2)}$~\cite{Stasi2023b}, or by allowing for operation at lower values of $|\alpha|^2$, and these should be considered in future experiments. While we exploit fusion-like operations~\cite{Browne2005} in this protocol, more recent approaches such as ``two-hierarchy entanglement swapping”~\cite{Xu2017} have also been shown to mitigate some of these effects. Applications requiring high efficiencies for heralding distributed entangled states, e.g., DIQKD~\cite{Kolodynski2020} and quantum repeaters~\cite{Sangouard2007}, will nonetheless need to leverage deterministic single-photon sources to significantly increase the non-postselected fidelities obtained here and improve the distribution rates. The coupling efficiency and performance of single-photon sources is now reaching a level where clear scaling advantages can be achieved~\cite{Tomm2021}, especially for multiphoton experiments~\cite{Cao2023}. Moreover, as we have shown, care must be taken to use high-extinction-ratio optical routing devices (with possible associated experimental redesign), without which the scheme is significantly limited in the regime of interest even for deterministic single-photon sources. 


\begin{acknowledgments}
We thank Laura Dos Santos Martins and Vasiliki Angelopoulou for useful discussions. This work was supported by the Swiss National Science Foundation SNSF (Grant No.~200020\_182664), the NCCR QSIT and the Swiss State Secretariat for Research and Innovation (SERI) (Contract No. UeM019-3). 

\medskip
F.J.M. and P.C. contributed equally to this work.
\end{acknowledgments}

\medskip
\appendix
\bibliographystyle{myabbrvnat}
\bibliography{references}

\end{document}